\documentclass[aip,
amsmath,amssymb,
reprint,]{revtex4-1}
\usepackage{graphicx}
\usepackage{dcolumn}
\usepackage{bm}
\usepackage[utf8]{inputenc}
\usepackage[T1]{fontenc}
\usepackage{mathptmx}
\DeclareMathOperator{\arcsinh}{arcsinh}

\usepackage[caption=false]{subfig}
\usepackage{color,soul}
\begin{document}
\preprint{AIP/123-QED}
\title{Effects of surface charge and environmental factors on the electrostatic interaction of fiber with virus-like particle: A case of coronavirus }

\author{D. N. Dung}
\email{dung.dinhngoc@phenikaa-uni.edu.vn,
anh.phanduc@phenikaa-uni.edu.vn}
\affiliation{Faculty of Materials Science and Engineering, Phenikaa University, Yen Nghia, Ha Dong, Ha Noi 12116, Vietnam}

\author{Anh D. Phan}
\affiliation{Faculty of Materials Science and Engineering, Phenikaa University, Yen Nghia, Ha Dong, Ha Noi 12116, Vietnam}
\email[Author to whom correspondence should be addressed:]{anh.phanduc@phenikaa-uni.edu.vn}
\affiliation{Phenikaa Institute for Advanced Study, Phenikaa University, Yen Nghia, Ha Dong, Ha Noi 12116, Vietnam}
\author{Toan T. Nguyen}
\affiliation{Key Laboratory for Multiscale Simulation of Complex Systems, and Faculty of Physics, VNU University of Science, 
Vietnam National University -- Hanoi,
 334 Nguyen Trai Street, Thanh Xuan District, Hanoi 10000, Vietnam}
 \author{Vu D. Lam}
\affiliation{Graduate University of Science and Technology, Vietnam Academy of Science and
Technology, 18 Hoang Quoc Viet, Hanoi, Vietnam}
\begin{abstract}
     \textbf{Abstract:}  We propose a theoretical model to elucidate intermolecular electrostatic interactions between a virus and a substrate. Our model treats the virus as a homogeneous particle having surface charge and the polymer fiber of the respirator as a charged plane. Electric potentials surrounding the virus and fiber are influenced by the surface charge distribution of the virus. We use Poisson-Boltzmann equations to calculate electric potentials. Then, Derjaguin's approximation and a linear superposition of the potential function are extended to determine the electrostatic force. In this work, we apply this model for coronavirus or SARS-CoV-2 case and numerical results quantitatively agree with prior simulation. We find that the influence of fiber's potential on the surface charge of the virus is important and is considered in interaction calculations to obtain better accuracy. The electrostatic interaction significantly decays with increasing separation distance, and this curve becomes steeper when adding more salt. Although the interaction force increases with heating, one can observe the repulsive-attractive transition when the environment is acidic.
\end{abstract}
\maketitle
\section{INTRODUCTION}
Intermolecular force is important to determine the stability and persistence of viruses on the surface substrate. Several approaches based on experiment have been performed to elucidate the interaction of respiratory viruses such as SARS-CoV-2 \cite{van2020aerosol,ACSAMI}, SARS \cite{van2020aerosol, PMID:14631830}, MERS \cite{MERS1,MERS2}, and Ebola \cite{Ebola} with many types of common inanimate substrates, including glass, copper, stainless steel, and cloth. Nevertheless, experimental methods need complex equipment to perform an accurate measurement. Safety issues in experiments are also serious and significantly slow down the speed of research. On the other hand, theoretical approaches allow us to have better insights with high accuracy overcoming the above-mentioned drawbacks.

Several theoretical models have been proposed to study the electrostatic properties of viruses. The soft colloid model of Ohshima \cite{OHSHIMA1994474,OHSHIMA201373,OHSHIMA1995189,ohshima1994electrostatic,bell1970approximate} describes bacteria and viruses as biocolloidal particles having a neutral core covered by an ion permeable polyelectrolyte layer. On the contrary, several studies of Nguyen et al. \cite{nguyen2011rna} and \v{S}iber et al.\cite{vsiber2007role, vsiber2012energies} proved that the viral capsid, paradigm of a soft particle, may bear a volumetric core charge. Inspired by this idea, Phan and co-workers \cite{phan} considered a virus as a core-shell system with a constant charged hard core and analyzed its potential profile. McDaniel et al. \cite{mcdaniel2015electrostatic} extended Phan's model to capture the pH dependence of shell charge density. L{\'o}pez-Garc{\'i}a et al. \cite{lopez2003suspended} later considered the ion partitioning effect and numerically investigated this phenomenon for a soft particle with a core bearing a surface charge. Ganjizade et al. \cite{ganjizade2017effect, ganjizade2018effect} derived analytical expressions to describe the electrostatics of a particle with a volumetrically charged core under influences of the ion partitioning. While assumptions of the uniform surface charge distribution in biological structures have been widely used, the response of charge distribution to the local electrostatic potential remains barely exploited although this is an important factor, particularly when the separation distance between two charged objects is sufficiently close. The emergence of an induced electric field causes charge redistribution and may consequently alter the interaction between them. Recently, several simulation works done by Podgornik and his co-workers \cite{advzic2014field,refer,bovzivc2017ph} took into account the inhomogeneities of virus surface charge and indicated the importance of charge regulation on the electrostatic properties of the virus. However, simulations require heavy computational workload and are time-consuming, especially for a virus with complicated surface geometry such as SARS-CoV-2. 

SARS-CoV-2 virus has caused an outbreak resulting in COVID-19 pandemic with many severe consequences \cite{WHO,WB}. Among many adopted strategies to counteract the infection of SARS-CoV-2, wearing masks is an effective way to limit the spread of the virus and protect ourselves \cite{:/content/10.2807/1560-7917.ES.2020.25.6.2000110,dbouk2020respiratory,kumar2020perspective}. Since the major pathway for transmission of virus occurs through the contact with the respiratory droplet\cite{smith2020aerosol,busco2020sneezing}, face masks, such as surgical masks and N95 respirators, are highly recommended protection equipment during the pandemic. Their layer structure consists of a nonwoven electrically charge polymer fiber to strengthen the efficiency of filtration performance, which leads to reduced transmission chance of respiratory droplets \cite{filter1,CHU20201973,dbouk2020coughing,verma2020visualizing}. Understanding the interaction force between the virus and inanimate substrate, such as fiber, is essential to study the filtration process \cite{tsai2002different,ardkapan2014filtration} and decontamination method\cite{doi:10.1021/acs.estlett.0c00534, doi:10.1021/acsnano.0c02250,OH2020116386}. 

In this work, using SARS-CoV-2 as a case to study,  we propose a simple theoretical approach for calculating the interaction of the  virus with the surface charge of the polymer fiber of the mask. We employ Poisson-Boltzmann equations to obtain electrostatic potential profiles surrounding objects and combine these potentials with Derjaguin's approximation to determine interaction forces. Since the surface charge density of particle-like viruses is affected by the local electrostatic field induced by fiber plane, the effects of the induced electric field on our calculations are considered. We also discuss influences of other environmental factors such as pH, salt concentration, and temperature on the interaction force to further verify our model. 

\section{THEORETICAL BACKGROUND}
In this section, we construct a theoretical description and parameters for the SARS-CoV-2 virus immersed in the electrolyte and polymer fiber of the mask. Figure \ref{fig:1} illustrates the model of the virus and the fiber. The virus is treated as a charged sphere, while one can consider the cylindrical fiber surface in its vicinity as a flat surface since the virus is so small compared to the fiber. The interaction between these species strongly depends on their potential profile. Therefore, to determine the interaction force, first one needs to know the electrostatic potential of these objects when they are isolated.
    \subsection{Isolated particle-like SARS-CoV-2 virus}
    
\begin{figure}
 \includegraphics[width=\linewidth]{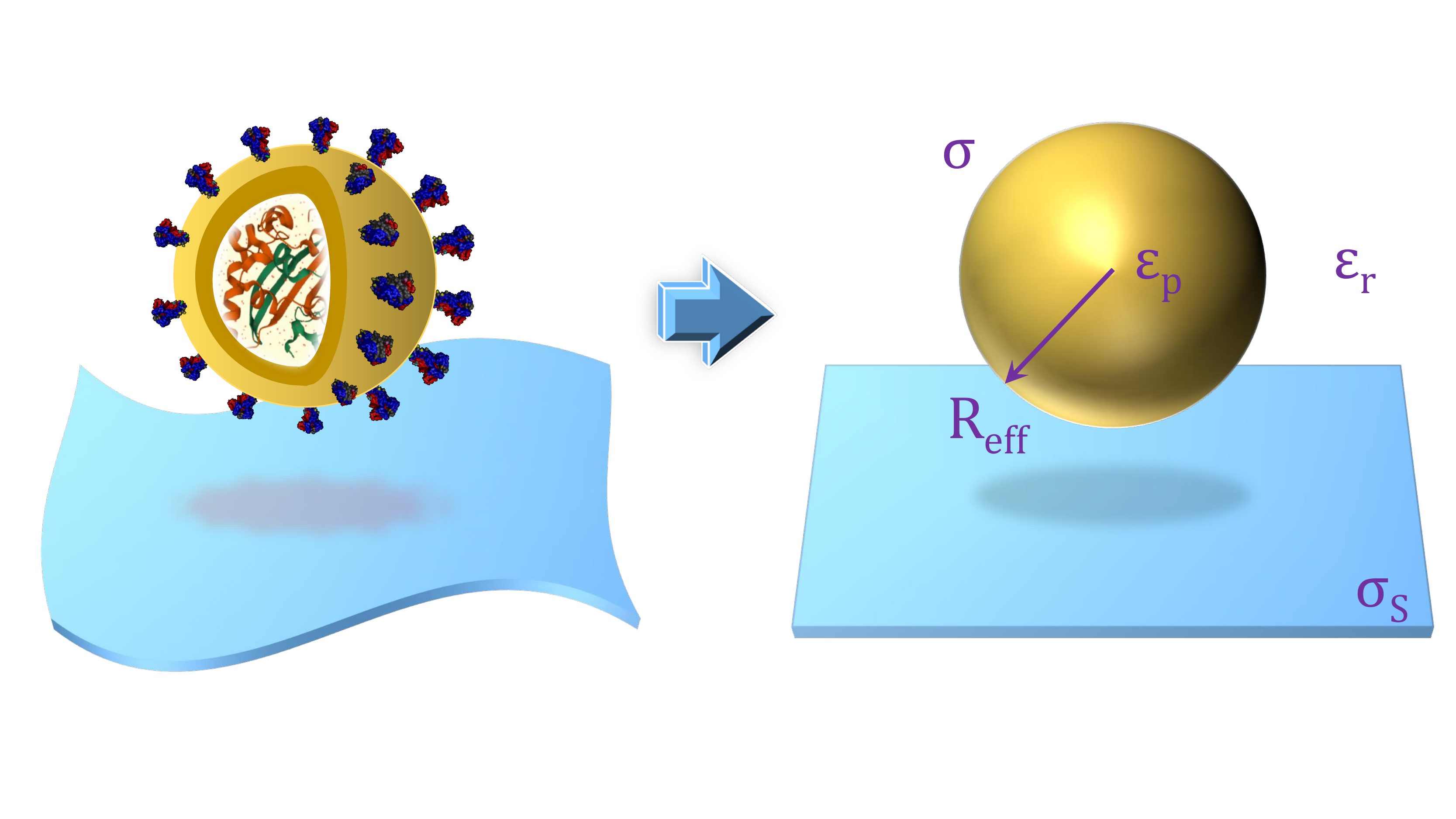}
 \caption{Illustration for the theoretical geometry of SARS-CoV-2 virus and planar fiber in an electrolyte solution.}
 \label{fig:1}
\end{figure}

A SAR-CoV-2 virus consists of a nucleocapsid enveloped by a spherical lipid membrane with S proteins attached on its surface \cite{yao2020molecular}. In our calculation, the DNA/RNA core and its viral lipid envelope are electrically screened, and only S proteins are responsible for the electrostatic interaction. These assumptions are consistent with a prior work \cite{heffron2021virus}. For S proteins, their shape and charge as described in \textbf{Fig. \ref{fig:1}} can be extracted from the Protein Data Bank entry PDB: 6VXX39 \cite{WALLS2020281}. For the sake of simplicity, S proteins are assumed to be perpendicular and uniformly distributed over the surface. Each S protein contains a large number and different types of Amino Acids (AAs). However, only AA residues, which are solvent accessible, are capable of contributing to the surface charge of the protein. These AAs can be classified based on their charging mechanism: ASP, GLU, TYR, and CYS  protonate to have negative charge, while ARG, LYS, and HIS types gain positive charge through deprotonation processes. Thus, a mathematical form of the surface charge of S protein is described by the Henderson-Hasselbalch equation \cite{bovzivc2017ph} 
\begin{align}
    q_0=&\sum_{i=ASP,GLU,TYR,CYS}\dfrac{ -e}{1+e^{- \ln 10(pH-pK_a^{(i)})}}\nonumber\\
    &+\sum_{i=ARG,LYS,HIS}\dfrac{ +e}{1+e^{ \ln 10(pH-pK_a^{(i)})}},
\end{align}
where  $e$ is the elementary charge and the sum takes over all types of AAs with $\pm$ corresponding to deprotonated (ASP, GLU, TYR, and CYS) and protonated (ARG, LYS, and HIS) AAs, respectively. $pK_a^{(i)}$ is the chemical dissociation constant of each type of AAs, and the value of $pK_a^{(i)}$ can be found in Ref.~\onlinecite{refer}. Upon analysis of SARS-CoV-2 by electron microscopy, the diameter of the particle-like virus ranges from 50 nm to 140 nm \cite{size1,menter2020postmortem, VARGA20201417}. In our model, a particle-like virus has the effective radius of $R_{eff} = 70$ $nm$, which is the distance from the center of the virus to S proteins. Because of the screening of the other charge, half of S protein charge $q'_0=q_0/2$ contributes to the interaction. The total of this charge of all S-proteins is assumed to spread over the virus surface, so it yields the surface charge of a particle-like virus with $N$ number of S proteins
\begin{equation}{\label{scharge}}
   \sigma=\dfrac{N q'_0 }{4\pi R_{eff}^2}.
\end{equation}

The electric potential is analytically described by the Poisson-Boltzmann equations associated with the corresponding boundary conditions for a particle-like virus\cite{PHAN201984}. These equations are
\begin{align}
        &\epsilon_p\epsilon_0\left(\dfrac{d^2\psi(r)}{dr^2}+\dfrac{2}{r}\dfrac{d\psi(r)}{dr}\right)=0 &\quad r<R_{eff}\nonumber\\
        &\epsilon_r\epsilon_0\left(\dfrac{d^2\psi(r)}{dr^2}+\dfrac{2}{r}\dfrac{d\psi(r)}{dr}\right)=2 n_0 e\sinh(\dfrac{e \psi}{k_BT}) &\quad r>R_{eff}\nonumber\\
        &\sigma (R_{eff})=\epsilon_p\epsilon_0\dfrac{d\psi}{dr}\bigg\vert_{r=R_{eff}^-}-\epsilon_r\epsilon_0\dfrac{d\psi}{dr}\bigg\vert_{r=R_{eff}^+}
\end{align}
  where $k_B$ is the Boltzmann constant, $T$ is the ambient temperature, $\psi$ is the potential of the isolated virus, $r$ is the radial distance to the center of a virus, $\epsilon_0$ is the vacuum permittivity, $\epsilon_r=78$ is the permittivity of the electrolyte, $\epsilon_p=4$ is the permittivity of the nucleocapsid,  $\kappa=\sqrt{\dfrac{2z_0^2e^2n}{\epsilon_r\epsilon_0 kT}}$ is the Debye-H{\"u}ckel parameter \cite{OHSHIMA201373}, and $z_0$ and $n$ are the ionic valence and ion number concentration in solution, respectively. In this work, we consider a monovalent electrolyte, and thus, $z_0=1$. In addition, the authors in Ref.\cite{refer} revealed that S proteins are mainly responsible for interactions of SAR-CoV-2 virus with the environment and effects of the structure of the virus interior can be ignored. For simplicity, we assume that the virus core is neutral.
    \begin{figure*}
{\includegraphics[width= 14cm]{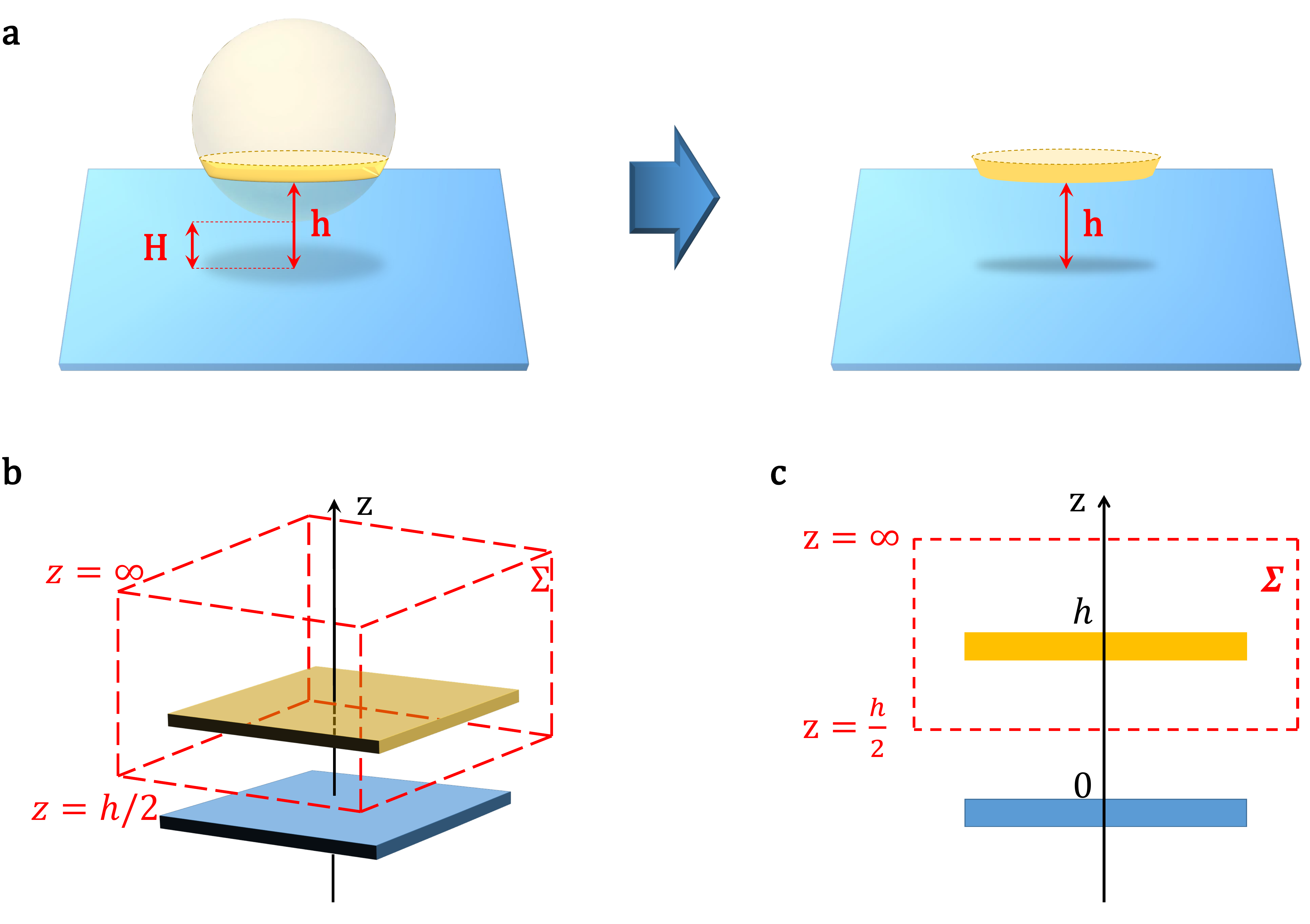}}\hfill
\caption{(a)  Derjaguin’s approximation for a spherical object interacting with the planar sphere, (b) and (c) 3D and 2D schematic of the enclosed surface $\Sigma$ according to the theory of  Verwey and Overbeek.  }
\end{figure*}     
The above equations have not yet been exactly solved. However, for the case of $\kappa R_{eff}\gg1$, an analytical expression of potential distribution around a spherical particle is approximately obtained by \cite{white1977approximate}
  \begin{equation}
      \psi(r)=\dfrac{2kT}{e}\ln\left(\dfrac{1+\gamma_{sp}\dfrac{R_{eff}}{r}e^{-\kappa(r-R_{eff})}}{1-\gamma_{sp}\dfrac{R_{eff}}{r}e^{-\kappa(r-R_{eff})}}\right),
  \end{equation}
   where $\gamma_{sp}=\tanh\left(\dfrac{e\psi_0}{4kT}\right)$, $\psi_0$ is the surface potential of virus, and the relation between surface charge and surface potential is
   \begin{equation}
       \sigma=\dfrac{2\epsilon_r\epsilon_0\kappa kT}{e}\left[\sinh\left(\dfrac{e\psi_0}{2kT}\right)+\left(\dfrac{2}{\kappa R_{eff}}\right)\tanh\left(\dfrac{e\psi_0}{2kT}\right)\right].
   \end{equation}
 
    \subsection{Isolated polymer fiber}
Since the dimension of the polymer fiber is much larger than the radius of virus \cite{fibsiz}, the polymer fiber can be considered as a semi-infinite plane as shown in \textbf{Fig. 1}. To the zeroth-order approximation, we assume that there is no presence of induced charges during the interaction with the virus and the surface charge of the fiber is fixed at $\sigma_S=-1e/nm^2$. The gradient of electrostatic potential near the fiber obeys the Poisson-Boltzmann equation
\begin{equation}\label{5}
    \epsilon_r\epsilon_0\dfrac{d^2\phi(z)}{dz^2}=2n_0 e \sinh\left(\dfrac{e\phi(z)}{kT}\right),
\end{equation}
where z-axis is perpendicular to the planar surface with its origin $z=0$ at the surface (as depicted in \textbf{Fig. 2}). When the virus comes to the proximity of the fiber surface, the potential of the plane $\phi (z)$ plays a role as local potential which can alter the surface charge distribution of the virus. This effect will be discussed further in the section III.C. The potential profile of the semi-infinite plane derived from solving \textbf{Eq. (\ref{5})} is
   \begin{equation}
       \phi(z)= \dfrac{2k_BT}{e}\ln\left(\dfrac{1+\gamma_{pl} e^{-\kappa z}}{1-\gamma_{pl} e^{-\kappa z}}\right),
   \end{equation}
   where $\gamma_{pl}=\tanh\left(\dfrac{e\phi_0}{4kT}\right)$ (where $\phi_0$ being the surface potential of the plane). An analytical expression of $\phi_0$ is
   \begin{equation}
      \phi_0=\dfrac{2k_BT}{e}\arcsinh\left(\dfrac{e\sigma_S}{2\epsilon_r\epsilon_0\kappa kT}\right).
   \end{equation}
\section{INTERACTION OF PLANAR FIBER WITH SARS-CoV-2 VIRUS}
In this section, we use Derjaguin's approximation and a linear superposition of potentials to calculate the interaction force between the SARS-CoV-2 virus and the electret fiber of the mask. The effect of electret fiber potential on the virus charge surface is considered.  We also delve into how the calculated force depends on the separation distance and environmental factors including pH, and ion concentration.

     \subsection{Derjaguin's approximation}
    
To calculate the interaction force, a spherical object is modeled as a superposition of infinitesimal parallel parts as shown in \textbf{Fig. 2(a)}. According to Derjaguin's approximation \cite{derjaguin1940repulsive}, in the limit of $\kappa R_{eff} \gg 1$, the interaction electrostatic potential per unit area of this part and planar surface is equal to the energy of interaction per unit area of two infinite planes at the same separation distance $h$ [\textbf{Fig. 2(b)}]. Thus, the interaction force between the sphere-like virus and planar fiber is estimated as
 \begin{equation}\label{int}
F=2\pi R_{eff}\int_H^{\infty} F_{pl}(h) dh,
\end{equation}
where $F_{pl}$ is the interaction force between the planar part of the spherical particle and substrate per unit area and $H$ is the distance from the virus surface to the fiber. To calculate the virus-fiber interaction, the explicit form of $F_{pl}$ must be known.

 When the charge body is immersed in an electrolyte, it exhibits the excess osmotic pressure $\Delta \Pi$ that originated from the electrical double layer and Maxwell forces. These forces are strongly dependent on the interactions of surface charge body and counterions in the electrolyte. Thus, the total interaction force of two planes can be obtained from the integration of the total stress tensor, which consists of the Maxwell electrostatic and Van't Hoff osmotic component, over the surface enclosing either one of the two planes $\Sigma$ \cite{KROZEL1992365},

\begin{align}\label{tot}
    F_{pl}(h)
    =&2 k_BT n_0\left[\cosh\left(\dfrac{e\varphi(h/2)}{kT}\right)-1\right]\nonumber\\
    &-\dfrac{1}{2}\epsilon_r\epsilon_0\left(\dfrac{d\varphi(z)}{dz}\bigg\vert_{z=h/2}\right)^2,
\end{align}
with $\varphi$ being the potential between two planes.
 
 In our calculations, the closed surface is chosen according to the theory of Verwey and Overbeek \cite{ohshima2011biophysical}. With this choice, the surface $\Sigma$ consists of two planes $z=-\infty$ and $z=-\dfrac{h}{2}$ as shown in \textbf{Figs. 2(b)}, and \textbf{2(c)}.  For simplicity, $\varphi(z)$ is the linear superposition of potentials of two isolated planes. In general, the surface charge of spherical virus can be influenced by the planar potential. Therefore, the value of $\gamma_{sp}$ of the plane divided from the spherical particle depends on the separation distance $h$. Based on \textbf{Eq. (7)}, one can derive the form of the potential
    \begin{equation}\label{psi}
        \varphi(z)=\dfrac{2k_BT}{e}\ln\left[\dfrac{(1+\gamma_{pl}e^{-\kappa z})(1+\gamma_{sp}(h)e^{-\kappa z})}{(1-\gamma_{pl}e^{-\kappa z})(1-\gamma_{sp}(h)e^{-\kappa z})}\right].
    \end{equation}
    
From \textbf{Eqs. (\ref{tot})} \textbf{(\ref{psi})}, we obtain  
\begin{align}\label{Fpl}
    F_{pl}(h)&=16 n_0 k_BT e^{-\kappa h}\dfrac{4\gamma_{pl}\gamma_{sp}(h)\left(1+ \gamma_{pl} \gamma_{sp}(h) e^{-\kappa h}\right)^2}{\left(1-\gamma_{pl}^2e^{-\kappa h}\right)^2\left(1-\gamma_{sp}(h)^2e^{-\kappa h}\right)^2} \nonumber\\
    &\approx 64 n_0 k_B T \gamma_{pl}\gamma_{sp}(h) e^{-\kappa h
    }.
\end{align}

\begin{figure}
{\includegraphics[width = \linewidth]{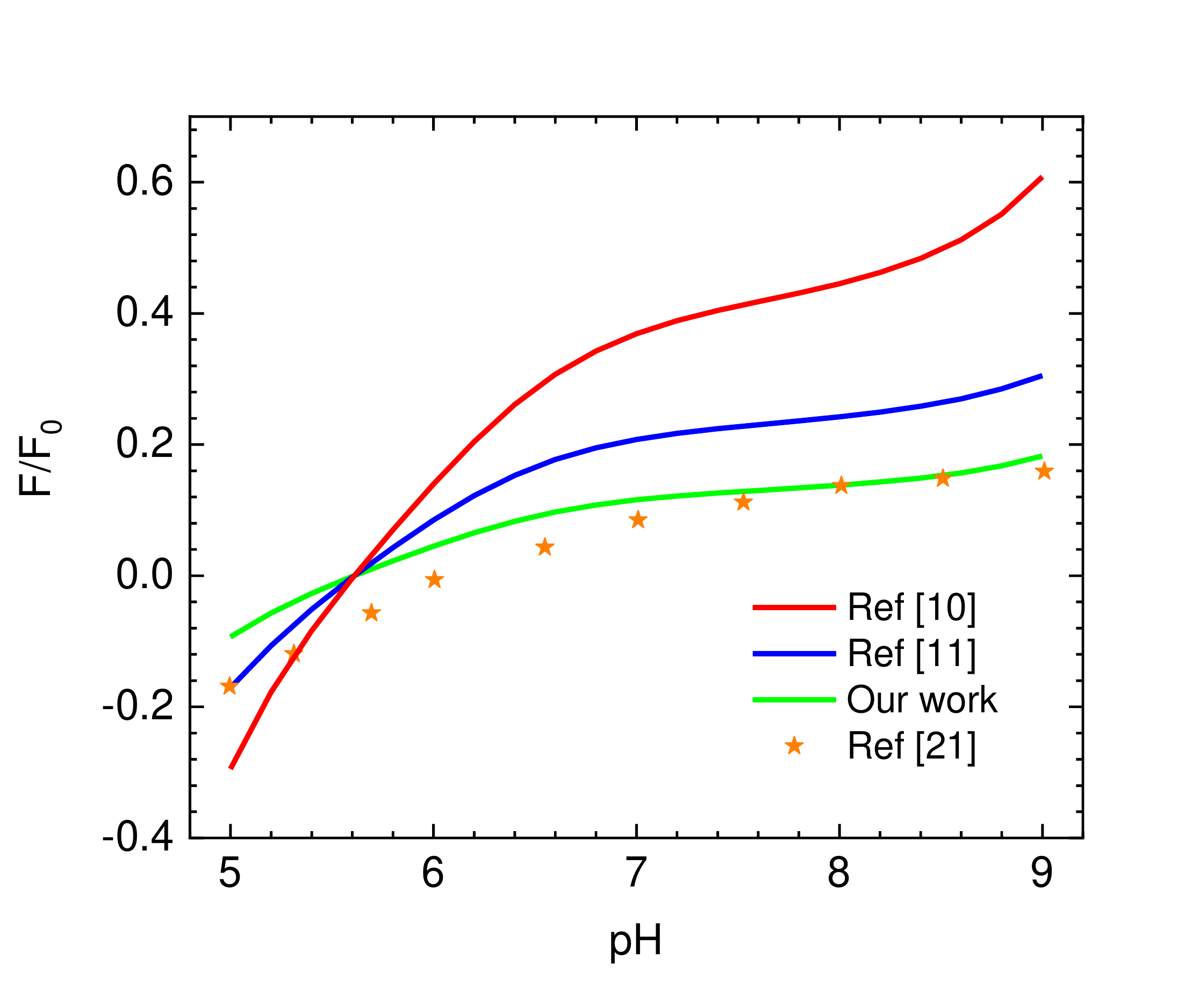}}
    \caption{The pH dependence of the scaled interaction forces between the particle-like SARS-CoV-2 virus with constant surface charge and the planar fiber}
    \end{figure}

\begin{figure}
{\includegraphics[width = \linewidth]{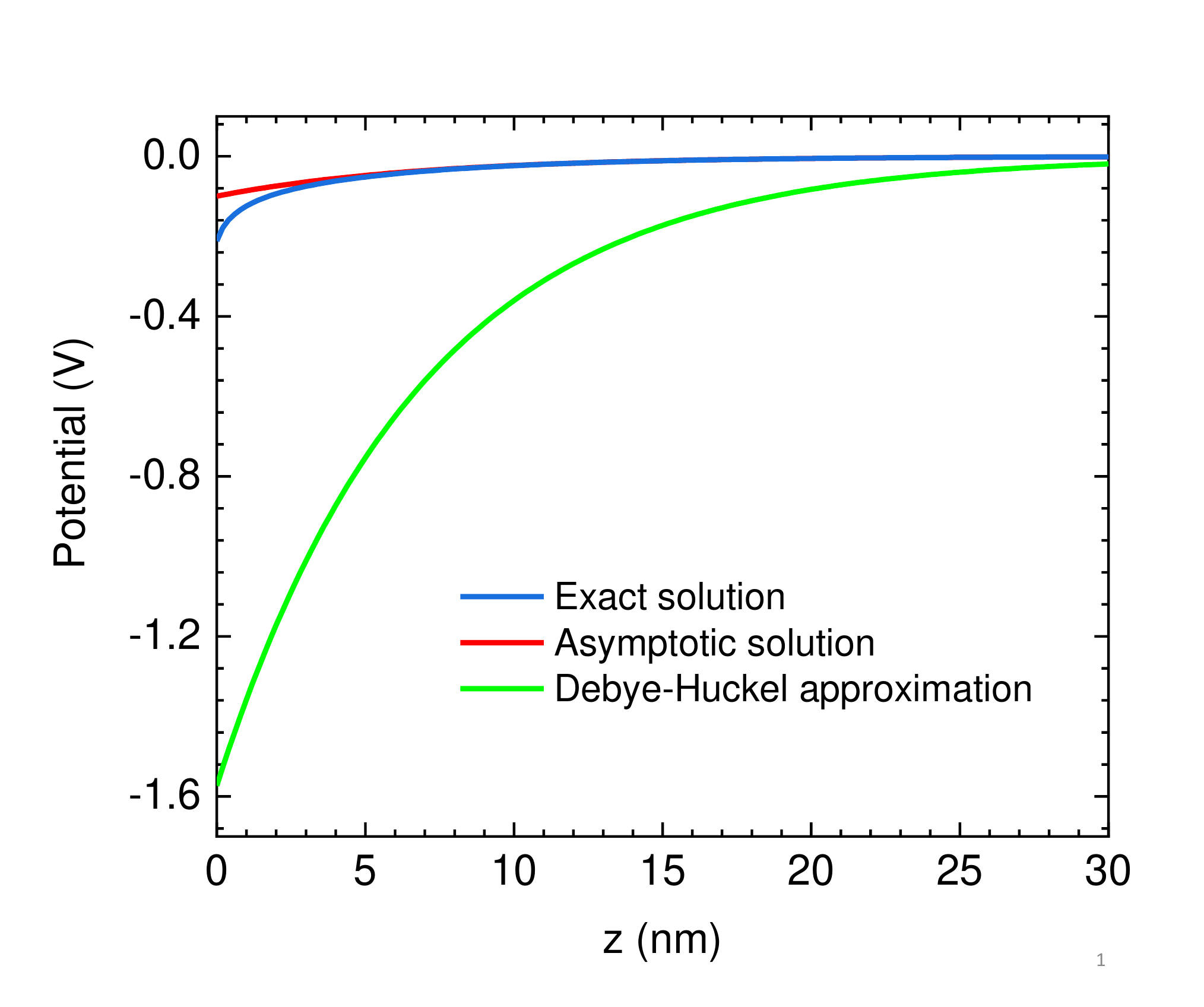}}
    \caption{Potential distribution around the charged plate derived from solving the Poisson-Boltzmann equation.} 
    \end{figure}    
\subsection{Particle-like virus with constant charge surface}       
First, we assume that when the virus reaches the vicinity of the planar fiber, the surface charge density of the virus is not influenced by the planar electrostatic field. In the absence of the local field induced by the fiber, according to \textbf{Eq. (2)}, at a particular value of $pH$, the surface of the particle-like virus is kept constant. By using \textbf{Eq. (\ref{Fpl})}, the scaled interaction force with $F_0=64\pi\epsilon_r\epsilon_0\kappa R_{eff}\left(\dfrac{k_B T}{e}\right)^2$ between the virus bears $N=60$ S-proteins on its surface and the fiber as a function of $pH$ at a salt concentration of $2$ mM, separation distance of $2.1$ nm and at $T=25$ $^0C$ is calculated and shown in \textbf{Fig. 3}. The $pH$ level in this work ranging from 5.6 to 8.4 is consistent with the regular $pH$ level of the respiratory droplet. In this approach, the force is reversed from attractive to repulsive at the isoelectric point of the charge surface with a $pH$ value of 5.6. These results suggest that when a respiratory droplet is attached to the mask with negative charge fiber, the virus within it is more readily to adhere to the substrate in an acidic environment.
    
In \textbf{Fig. 3}, we also compare our numerical results with some previous works that are also based on the Derjaguin-Landau-Verwey-Overbeek (DLVO) method \cite{ohshima1994electrostatic,bell1970approximate} and a recent simulation \cite{refer}. It is clear that previous models do not work well in this case since they overestimate the magnitude of interaction force. The main reasons for the deviation are as follows. First, the effects of the surface charge distribution on the electrostatic interaction were not captured. Second, the models in Refs. [\onlinecite{ohshima1994electrostatic}] and [\onlinecite{bell1970approximate}] are constructed from the solution of the linear Poisson–Boltzmann equation based on the Debye–H{\"u}ckel approximation for the planar potential. Our Eq. (3) is significantly simplified and these calculations are validated at a very small potential. To elucidate this, in \textbf{Fig. 4} we compare the planar potential derived from the exact solution, the Debye–H{\"u}ckel approximation solution, and the asymptotic solution of the Poisson–Boltzmann equation. These potentials have the following form
\begin{align}
    \phi_{DH}(z)=\dfrac{\sigma_s}{\epsilon_r\epsilon_0\kappa}e^{-\kappa z},\\
    \phi_A (z)=\dfrac{4kT}{e}\gamma_{pl}e^{-\kappa z}.
\end{align} \textbf{Fig. 4} shows that the potential profiles derived from the Debye–H{\"u}ckel approximation overestimate the exact solution. Since the Debye–H{\"u}ckel approximation is only valid for the case of low potential, this deviation indicates that the potential in this case is no longer small to be treated by the Debye–H{\"u}ckel approximation. While the asymptotic form describes the potential distribution around the planar well at a large separation distance. On the other hand, \textbf{Eq. (12)} employs the exact solution of the nonlinear Poisson–Boltzmann equation, so its results provide better agreement with finite element simulation than calculations in Refs. [\onlinecite{ohshima1994electrostatic}] and [\onlinecite{bell1970approximate}]. Even though, the theoretical curves of our model qualitatively describe the variation of simulation data, their deviations are relatively large. Consequently, it is essential to consider the effects of fiber potential on the virus surface charge rather than the condition of constant charge surface.
\begin{figure}
{\includegraphics[width = \linewidth]{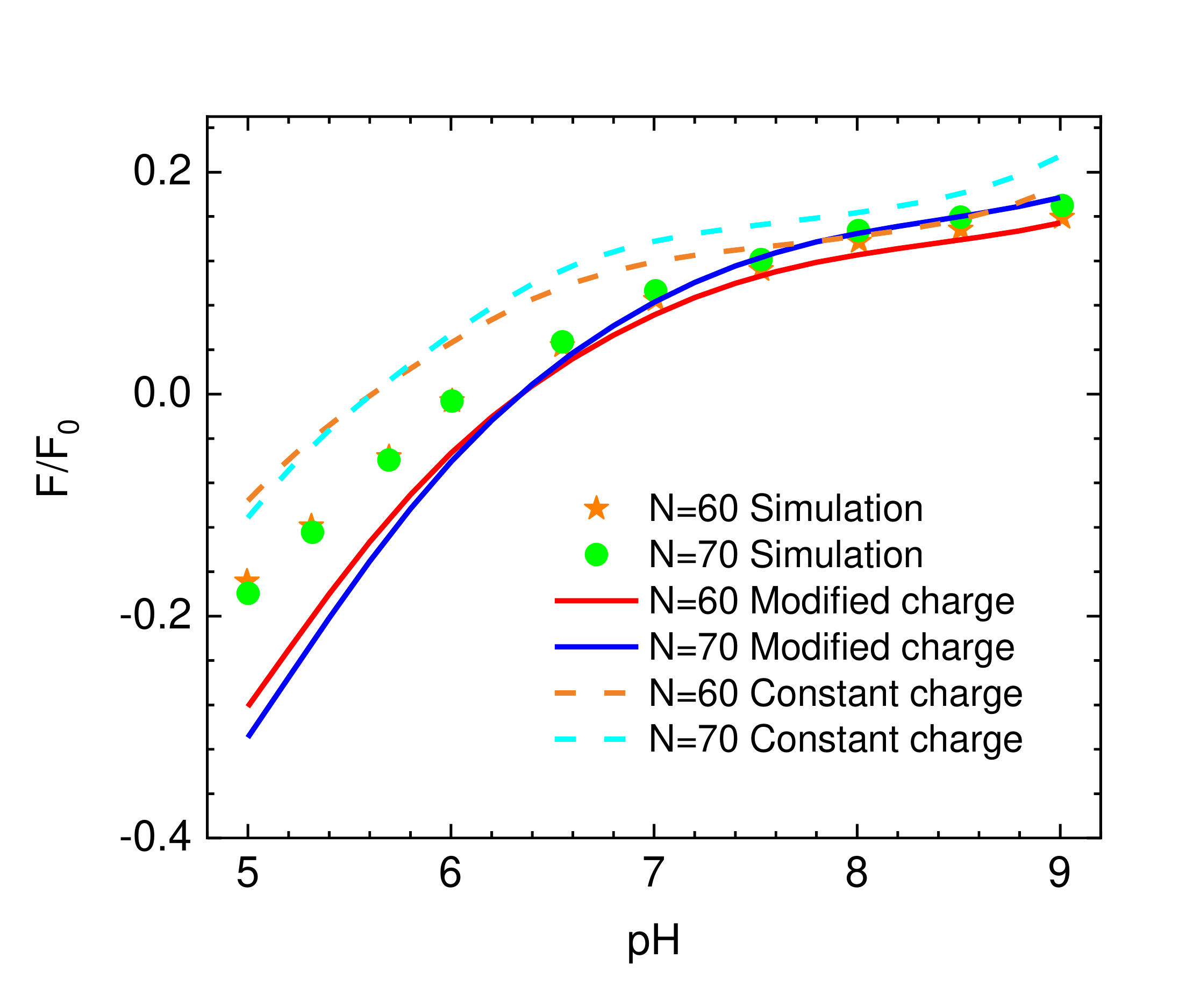}}
    \caption{Scaled interaction forces in the case with (solid line) and without (dashed line) charge regulation as a function of pH at $n = 2$ $mM$, $H-2.1$ nm and $T=25^oC$. The symbols indicates the simulation results.} 
    \end{figure}  

  \begin{figure*}[htp]
     \includegraphics[width = 16cm]{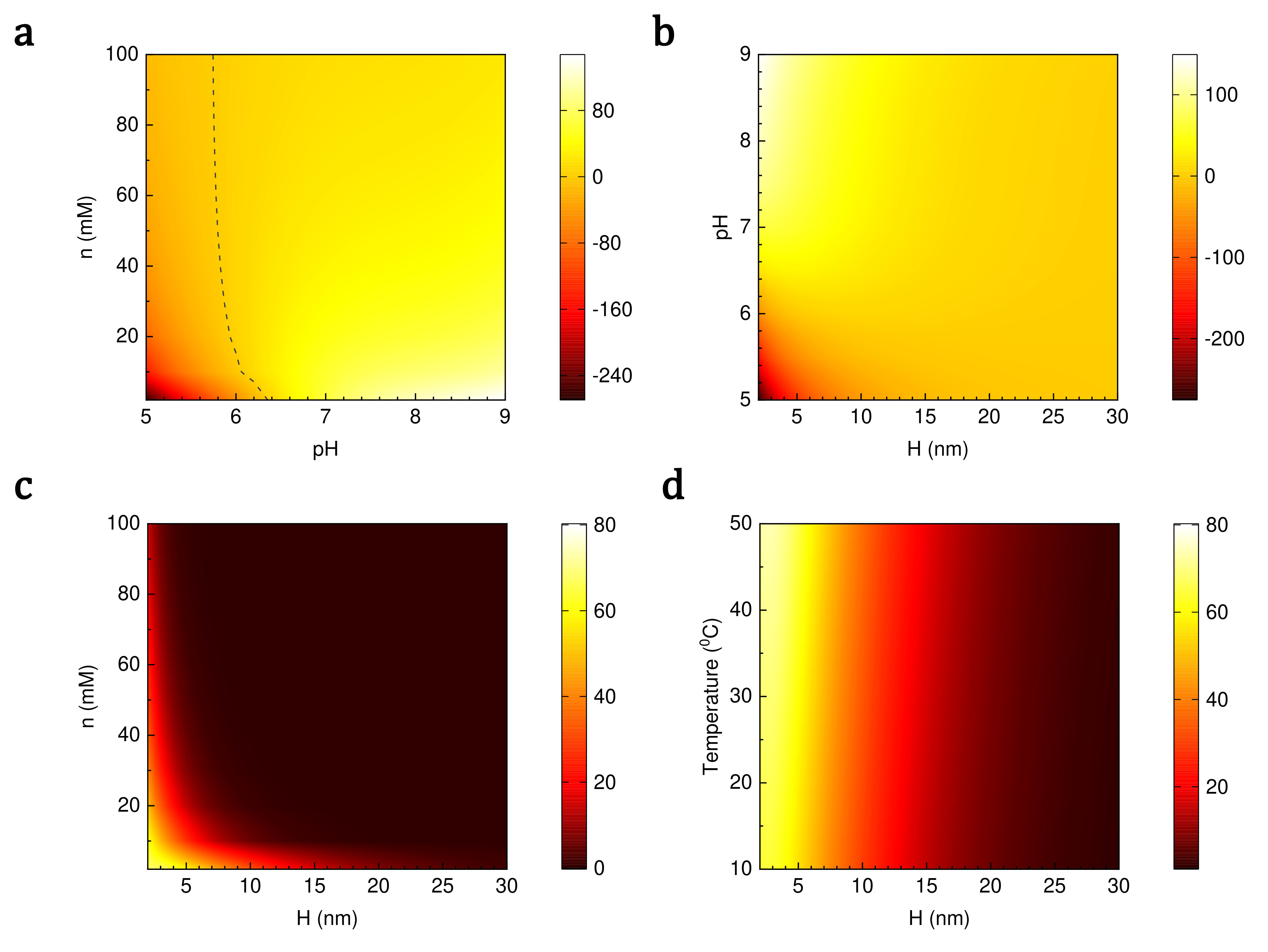}
    \caption{Contour plots of interaction forces exerted on the virus with 60 S protein in the unit of pico Newton (pN) between (a) $n$ and pH for $H=2.1 nm$ and $T=25^oC$ (b) pH and $H$ for $n=2 mM$ and $T=25^oC$; (c) $n$ and $H$ for $pH=7$ and $T=25^oC$; and (d) $T$ and $H$ for $n=2 mM$ and $pH=7$.} 
    \end{figure*}
\subsection{Particle-like virus with modified charge surface}
To improve the accuracy of our calculations, we take into account the effects of the electrostatic field of the polymer fiber on the charge surface density of the particle-like virus. The roles of the induced field become more important when the virus is closer to the fiber. The local electrostatic potential of the fiber causes the shift of $pK_a^{(i)}$ values: $pK_a^{(i)} \rightarrow pK_a^{(i)}- e\phi(h)/\ln 10 k_B T$. Consequently, the surface charge of the virus is redistributed as a function of the separation distance. Using the modified Henderson-Hasselbalch equation gives us the distribution of the modified charge density in the surface of the particle-like virus \cite{bovzivc2017ph, CR}

\begin{align}
    \sigma(h)=&\sum_{i=ASP,GLU,TYR,CYS}\dfrac{ -Ne/8\pi R_{eff}^2}{1+e^{- \ln 10(pH-pK_a^{(i)})-e\phi(h)/k_B T}}\nonumber\\
    &+\sum_{i=ARG,LYS,HIS}\dfrac{ Ne/8\pi R_{eff}^2}{1+e^{ \ln 10(pH-pK_a^{(i)})+e\phi(h)/k_B T}}.
\end{align}
Unlike the case of constant charge surface, the charge distribution at different points on the surface of the virus, in this case, depends on the distance $h$ from the tagged point to the planar. By taking integration over $h$ in \textbf{Eq. (\ref{int})}, we can determine the interaction force between a sphere-like SARS-CoV-2 virus and an infinite plane.

\textbf{Figure 5} shows the electrostatic interaction between the SARS-CoV-2 virus having different numbers of S proteins and the mask as a function of $pH$ at the ion concentration of $2$ mM, separation distance of $2.1$ nm and the temperature of $25 ^o$C. One can observe a quantitative good agreement between our calculations and simulation \cite{refer}. Upon the increase in the number of S proteins, the interaction force magnitude is always enhanced. It is inline with \textbf{Eq. (15)} since the surface charge density of the virus is proportional to the number of S proteins. Additionally, the interaction forces reverse from attraction to repulsion at $pH \approx 6.1$. Compared to the case of the constant virus charge surface, the transition point is shifted toward the basic $pH$ region. Therefore, when the virus approaches the mask, under the local electrostatic of the electret fiber, not only the charge distribution but also the overall charge surface is modified depending on the fiber potential. \textbf{Fig. 6(a)} shows the dependence of interaction force on pH and salt concentration with the transition point represented by the dashed line. The curve of this line indicates that when the salt concentration increases, the transition point shifts to the low value of pH and gradually approaches the value of the transition point in the case of constant surface charge. This result is in agreement with previous works since the increase in the salt concentration leads to weakening of the potential of the fiber. Thus, the effect of fiber potential on the surface charge of the virus decreases and becomes negligible when the salt concentration is very high.
    
To further validate our analytical model, we also investigate the dependence of the calculated interaction force on environmental parameters. In \textbf{Fig. 6(b)}, we illustrated the impact of the separation distance on the interaction force between particle-like virus and the planar fibre. At the temperature of $25$ $^o$C and the salt concentration of $2$ mM, the interaction force decays with increasing separation distance at all values of $pH$. It is reasonable since the potential of the particle-like virus and the plane gradually approach zero at a large distance as one can see from \textbf{Eqs. (4)} and \textbf{(7)}. From \textbf{Eq. (15)}, this result also means that the electric potential of the polymer fiber has a huge impact on the virus charge surface when the distance between them is shortened. In additional to an effect of the separation distance, a substantial effect of the salt concentration on the interaction force for $N=60$, $pH=7$, and $T=25^o$C can be observed in \textbf{Fig. 6(c)}. At the same pH, the interaction decays faster with an increase in the salt concentration. For $n=100$ mM, the force reduces to near zero at $H\approx 5$ nm. This value for $n=10$ and 2 mM is $H\approx 15$ nm and  $H\approx 20$ nm, respectively. This variation originates from the increase in Debye-H{\"u}ckel parameter value $\kappa$ when the salt concentration increases. Upon this condition, the electric double layer thickness around each object decreases and it leads to the weakening of the screening effect. \textbf{Figure 6(c)} also confirms the dependence of interaction force on the separation distance which is mentioned above. Additionally, it is worth noting that since the droplet sticking in the respiratory often exposes to the surrounding medium, the variation in relative air humidity influences the evaporation rate \cite{humid} and consequently alters the salt concentration of the respiratory droplet. Thus, the humidity affects the interaction force and the stability of virion within the droplet.

In additional to humidity, temperature has a significant effect on the virion in the respiratory droplet. A recent work of Dbouk and Drikakis \cite{dbouk2020weather} indicated that high temperature and low relative humidity lead to a remarkable reduction of virus viability. Thus, to observe the effect of temperature on interaction force, we calculate the interaction force of the virus and the fiber in 2 mM solution at different temperatures. The numerical results for a temperature regime ranging from $10$ to $50$ $^o$C are shown in \textbf{Fig. 6(d)}. The main feature underlined by the results of this figure is that the magnitude of interaction force slightly increase with heating. This trend is also true when the virus completely binds to the surface of the mask. In this case, in additional to the long-range electric double layer force, the virus also exhibits the short-range hydrophobic interaction. \cite{PRIVALOV1988191} Upon the heating process, the strength of hydrophobic interaction increases due to its entropy-driven mechanism,\cite{stock2015direct,ACSAMI} which implies the stronger adhesion and persistence of the virus on the substrate. Even though the temperature can amplify the stickiness between the SARS-CoV-2 virus and the fiber, thermal treatment, such as dry heating is still capable of decontaminating viral pathogens on the respirator surface. There are two reasons to explain this mechanism. First, the temperature-induced denaturation leads to the destruction of the protein spike and virus at high temperatures. \cite{seifer2021thermal,kampf2020inactivation} Second, at extremely high temperatures, a reduction of hydrophobic interaction causes fewer effects on the virus. 

\section{CONCLUSION}
We have proposed our simple calculation based on the DLVO method for the interaction force of the SARS-CoV-2 virus and the electret fiber of the mask. The fiber is treated as an infinite plane while the virus is simplified to a spherical particle with charge of all S proteins smearing uniformly over the particle surface. We also compare our approach with other simple analytical models to unravel the importance of the exact solution of the Poisson-Boltzmann equation as well as the effect of fiber electric potential on the virus charge surface to determine the sign and magnitude of electrostatic interaction. Numerical results calculated by our model show the substantial dependence of interaction force on the surface charge of both object and environmental parameters that include pH, salt concentration, and temperature. The good agreement of these results and the prior work prove the high accuracy of our model. Therefore, with the simple yet precise tool that is established in this work, one can quantitatively calculate the interaction between the virus and the inanimate surface substrates and have better insight into this intermolecular interaction.

There are two main limitations in our calculations. First, we theoretically treat the SARs-CoV-2 virus as a spherical particle, while S proteins on the surface of the virus make the surface rough. The roughness could lead to the difference between our results and simulation. Second, our model is based on Derjaguin's approximation, which provides better results when $\kappa R_{eff}>10$. For $\kappa R_{eff}<5$, theoretical predictions can become less accurate. 

\textbf{AUTHOR DECLARATIONS}

\textbf{Conflict of Interest}

The authors declare that they have no conflict of interest.
\begin{acknowledgments}
This research is funded by Vietnam National University grant number QG.20.82.
\end{acknowledgments}

\textbf{DATA AVAILABILITY}

The data that support the findings of this study are available from the corresponding author upon reasonable request.

\nocite{*}
\bibliography{Bib}

\end{document}